# Hybrid III-V/SiGe solar cells on Si substrates and porous Si substrates

Pablo Caño[a], Manuel Hinojosa[a], Luis Cifuentes[a], Huy Nguyen[b], Aled Morgan[b],
David Fuertes Marrón[a], Iván García[a], Andrew Johnson[b] and Ignacio Rey-Stolle[a]

[a] Instituto de Energía Solar, ETSI de Telecomunicación,
Universidad Politécnica de Madrid, 28040 Madrid, Spain
[b] IQE plc, Pascal Cl, St. Mellons, Cardiff, CF3 0LW, United Kingdom

*Abstract* — **A tandem GaAsP/SiGe solar cell has been developed employing group-IV reverse buffer layers grown on silicon substrates with a subsurface porous layer. Reverse buffer layers facilitate a reduction in the threading dislocation density with limited thicknesses, but ease the appearance of cracks, as observed in previous designs grown on regular Si substrates. In this new design, a porous silicon layer has been incorporated close to the substrate surface. The ductility of this layer helps repress the propagation of cracks, diminishing the problems of low shunt resistance and thus improving solar cell performance. The first results of this new architecture are presented here.**

## I. Introduction

Among the wealth of approaches to combine III–V materials and silicon for PV applications, monolithic structures are of particular interest for their straightforward integration in current manufacturing lines. In this monolithic integration III-V and group-IV compounds are grown heteroepitaxially on a single substrate to produce a multijunction solar cell, which is then processed and integrated in an almost conventional way. However, the direct growth of III–V and group-IV semiconductors on silicon has to tackle key difficulties such as the large differences in lattice parameters and thermal expansion coefficients. Therefore, a smart engineering of the buffer layer between the silicon and the III–Vs is utterly instrumental in order to accommodate these dissimilar parameters.

These III-V/Si architectures have also been developed for lasers [1], bipolar transistors [2], photodetectors [3] or light emitting diodes [4]. Focusing on solar cells, several monolithic III-V/Si photovoltaic devices have been reported such as GaAsP/Si dual-junctions using GaP buffer layers [5], GaInP/GaAs solar cells on Ge/Si virtual substrates [6] or GaAsP/SiGe dual junctions grown on inactive Si substrates [7], [8]. In the latter design, depending on the composition of the $Si_{1-x}Ge_x$ alloy, the extra degree of freedom of bottom cell bandgap tunability is provided. Thus, this design allows to approach the optimum bandgap combination for a dual-junction solar cell [9]. In fact, GaAsP/SiGe tandem cells on Si have reached efficiencies over 20% [8], [10], [11]. In these structures, a $Si_{1-x}Ge_x$ graded buffer is grown on the silicon substrate varying its composition until a certain Ge content is reached [10], [12], [13]. The thickness of such buffer layers is typically between 5-15 $\mu$m in order to reduce the defect density and achieve the target bandgap (i.e. lattice parameter) while ensuring enough quality in the surface morphology [1], [7], [10], [13]. Accordingly, these kind of graded buffer layers are known as forward buffers.

In this context, an innovative architecture for the integration of III–V compounds and SiGe alloys on silicon substrates is presented in this work, by means of the so-called reverse buffers [14], [15]. This III–V/SiGe/Si approach is ambitious in terms of growth complexity since a germanium layer is directly deposited on the silicon substrate, despite the large lattice mismatch. Then, a subsequent $Si_{1-x}Ge_x$ graded buffer layer is grown, changing its composition from pure Ge to the desired final target (Ge ~75%). Because of the reduction of the lattice constant with the progressive increase in silicon content, reverse buffers are subjected to tensile strain. The tensile strain facilitates the movement of threading dislocations [14], [15], aiding their annihilation and therefore reducing the threading dislocation density. Such dislocation glide in less in forward buffers, which are subjected to compressive stresses. Reverse buffers can be made thinner due to the need of smaller variations in composition to reach the desired bandgaps (~1 eV), which in turn entails a decrease in the costs [14], [16]. Furthermore, the tensile strain in these layers produces a smoother surface than those grown under compressive strain [15], [17]. On the other hand, some morphological issues have been reported in these architectures, as the appearance of cracks [18] which cause a degradation in the device performance as we could observe and study in our previous design [19].

To address this issue, in this architecture, a porous silicon layer on the substrate subsurface has been implemented. Due to its higher ductility and flexibility as compared to other semiconductors [20], the porous layer is used to absorb strain, enhancing relaxation of the Ge nucleation [21], [22] and thus, reduce the probability of crack formation in the structure during the thermal ramps. Although these buffer layers and structures are still under development, the first cohort of tandem GaAsP/SiGe solar cells grown on silicon substrates with porous layers have been manufactured and characterized showing an enhancement in performance in relation to previous designs [19] with promising results.

| | Layer | Doping | Thickness |
|---|---|---|---|
| MBE | GaAs Contact Layer | n-1e19 cm$^{-3}$ | 500 nm |
| | AlInP Window | n-1e19 cm$^{-3}$ | 40 nm |
| | GaAs$_{0.8}$P$_{0.2}$ Emitter | n-1e18 cm$^{-3}$ | 100 nm |
| | GaAs$_{0.8}$P$_{0.2}$ Base | p-2e17 cm$^{-3}$ | 2 μm |
| | GaAs$_{0.8}$P$_{0.2}$ BSF | p-4e18 cm$^{-3}$ | 30 nm |
| | GaAs Tunnel Junction | | 40 nm |
| | Ga$_{0.6}$In$_{0.4}$P Nucleation | n-4e18 cm$^{-3}$ | 100 nm |
| CVD | Ge cap | | 5 nm |
| | Si$_{0.24}$Ge$_{0.76}$ Emitter | n-1e19 cm$^{-3}$ | 250 nm |
| | Si$_{0.24}$Ge$_{0.76}$ Base | p-1e15 cm$^{-3}$ | 6 μm |
| | Si$_{0.24}$Ge$_{0.76}$ BSF | p-4e18 cm$^{-3}$ | 150 nm |
| | SiGe Graded Buffer | p-4e18 cm$^{-3}$ | 2 μm |
| | Ge buffer | p-4e18 cm$^{-3}$ | 3 μm |
| | Porous Si | | |
| | p-Silicon substrate | | |

Fig. 1. Structure of the GaAsP/SiGe tandem with porous silicon buffer layer with thicknesses and doping levels.

## II. EXPERIMENTAL

GaAsP/SiGe solar cell samples were grown following a combined approach using both chemical vapor deposition (CVD) and molecular beam epitaxy (MBE). Firstly, porous silicon was created by means of electrochemical etching on 6-inch (100) 6° off towards [110] silicon wafers with a resistivity of 0.01 Ω·cm. Then, all group-IV layers were grown in an ASM Epsilon LPCVD reactor at a temperature of ~650ºC using standard precursors (SiH$_4$, GeH$_4$). A germanium nucleation layer was deposited directly on the silicon wafer, then a SiGe reverse graded buffer was grown changing its composition gradually from pure Ge to a 76% of germanium content. Afterwards, the Si$_{0.24}$Ge$_{0.76}$ bottom cell with E$_g$ ~1 eV was grown, and the group-IV part of the structure was then capped with a 5 nm Ge layer in order to prevent oxidation. Samples were then transferred to a Veeco Gen2000 MBE reactor, where the III-V part of the structure was completed. All III-V layers were grown at ~650ºC lattice matched to the Si$_{0.24}$Ge$_{0.76}$ bottom cell, forming the tunnel junction and the GaAs$_{0.8}$P$_{0.2}$ top cell (E$_g$ ~1.7 eV) [8], [23], [24]. This structure is depicted in Fig. 1 with the corresponding doping levels and thicknesses.

The visual inspection of the epi-wafers was conducted with a Leica DM 1750 M Nomarski microscope. Atomic Force Microscopy (AFM) images were generated with a multimode Nanoscope III A tool from Bruker in tapping mode. The reciprocal space maps of the structures were taken with a high-resolution X-ray diffraction, using a Panalytical X'Pert Pro MRD tool. A JSM-6335F Scanning Electron Microscope (SEM) was utilized to perform cross section analyses using both backscattered and secondary electron detectors. The free carrier concentration was measured by Electrochemical CV profiling (using a Dage CPV21 tool), whereas Secondary Ion Mass Spectroscopy (SIMS) was used to measure chemical concentration.

The epi-wafers were processed into 0.09 cm$^2$ (3 mm x 3 mm) solar cells with conventional photolithography techniques. As group-IV and III-V semiconductors present different processing requirements, both front and rear contacts were specifically designed to reduce thermal budgets. In this way, the high temperatures of conventional silicon metallizations were avoided, minimizing the risk of crack propagation and guaranteeing material compatibility. The front contact was made using electroplated gold (~600 nm thick) without any alloying. The rear contact was deposited with Electron Beam Physical Vapor Deposition (EBPVD) and consisted of a stack of Pd(50nm) / Ti(50nm) / Pd(50nm) / Al(1000nm) alloyed at 170ºC during 600 s. No antireflection coating (ARC) was deposited on the cells.

External Quantum Efficiency (EQE) was measured using a custom-made tool based on a 1000W Xe lamp, a triple-grating monochromator (JOBIN-YVON) and a lock-in amplifier. Spectral Photovoltage (SPV) measurements were performed in two-probe contact mode at room temperature using a 200W QTH-lamp filtered through a 1/8 monochromator (Oriel) as probe beam and a lock-in amplifier (Stanford). I-V curves were measured using the four-probe method with a Keithley 2602 source-meter instrument and a home-made AAA solar simulator based on a 1000-W Xe-lamp and an ORIEL 68820 stabilized power supply.

## III. RESULTS AND DISCUSSION

### A. Material characterization

In Figs. 2a and 2b we show the X-ray reciprocal space maps of structures grown on the porous Si substrates. Fig.2a corresponds to the group-IV layers only while Fig. 2b includes the whole structure. In these maps, the silicon substrate peak is clearly visible in the upper left part of both graphs and is taken as the origin of the angle coordinates. The Ge buffer peak can be observed with a mismatch of 4.36% (obtained using additional asymmeytric scans not shown). Then, the SiGe graded buffer layer spreads from the Ge peak until the SiGe bottom cell peak with a significant tilt. The new peaks that appear in Fig 2.b correspond to the III-V semiconductor layers (GaAs contact layer and GaAsP top cell). In general, they show a lower crystallographic quality since they cover a wider area.

The growth on porous silicon did not suppress totally the formation of fractures since cracks are still visible on the epi-wafer surface to the naked eye. Anyhow, the crack density has been significantly reduced as compared to conventional Si wafers [19]. The cracks were studied cross-sectional SEM and these observations suggest that they cross the complete structure, i.e. they run from the topmost GaAs contact layer, downwards to partially penetrate into the Si substrate, as can be seen in Fig. 3.

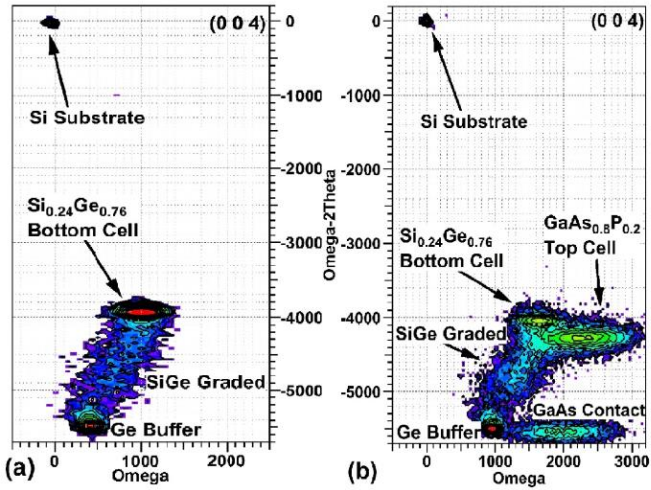

Fig. 2. a) Reciprocal space map (RSM) of the group IV part of the epi-wafer with porous silicon. b) RSM of the complete GaAsP/SiGe structure of the epi-wafer with porous silicon

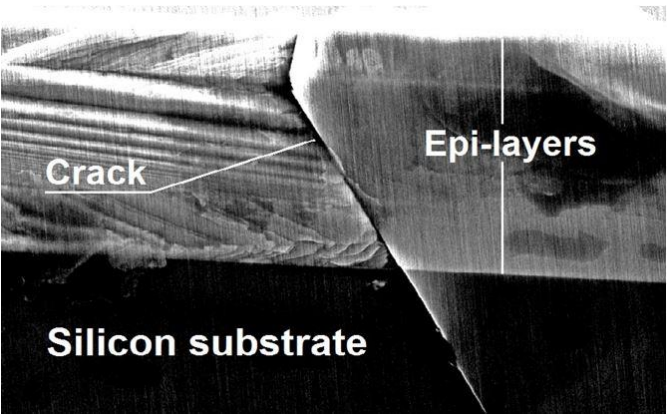

Fig. 3. Cross-section SEM image of the structure grown on conventional Si substrates (no porous layers) including a crack crossing the epi-layers and penetrating the Si wafer.

To provide deeper insight into the samples, Secondary-Ion Mass Spectroscopy (SIMS) and Electrochemical Capacitance Voltage (ECV) were measured in the SiGe bottom cell base in order to obtain dopant and carrier concentration versus depth, respectively. As can be observed in Fig. 4, the boron concentration shows the expected profile (constant at ~$10^{15}$ cm$^{-3}$), since a moderate doping level was designed in order to widen the space charge region and benefit from field-aided collection. On the other hand, it can be noticed that the free carrier concentration measured by ECV is much higher. This could be caused by intrinsic defects acting as efficient electron traps in the SiGe alloy [25]. Such a high doping will be detrimental for minority carrier collection in the bottom cell base.

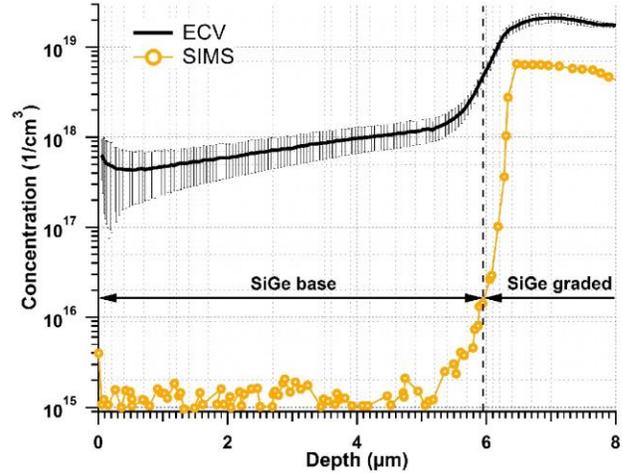

Fig. 4. Boron concentration measured by SIMS (in yellow) and carrier concentration measured by ECV (in black) of the group IV part.

*B. Solar cell results*

External Quantum Efficiency (EQE) was measured and can be seen in Fig. 5. The design on conventional Si substrates is represented in blue, whereas the new design with a porous silicon layer is depicted in red. In the previous design a functional GaAsP/SiGe tandem solar cell on silicon was demonstrated. Regarding the new design, the spectral response of the top cell has been improved, which indicates an enhancement of the crystallographic quality of the III-V material. This is probably due to a more effective relaxation of the group-IV layers, which leads to smoother surfaces and better subsequent growth of the III-V compounds [26]. However, we have not been able to measure the SiGe bottom cell EQE in samples grown on substrates with porous Si layers. The lack of response in the bottom subcell in a multijunction solar cell is a phenomenon widely reported in conventional triple-junction GaInP/Ga(In)As/Ge solar cells [27]. In these devices, it is frequent not to be able to measure the response of the Ge bottom cell as a result of a low shunt resistance, a low breakdown voltage, a high luminescent coupling with upper subcells or a combination of the above [27]. In our case there is a low shunt resistance in the SiGe subcell and the role of the other factors is still under study.

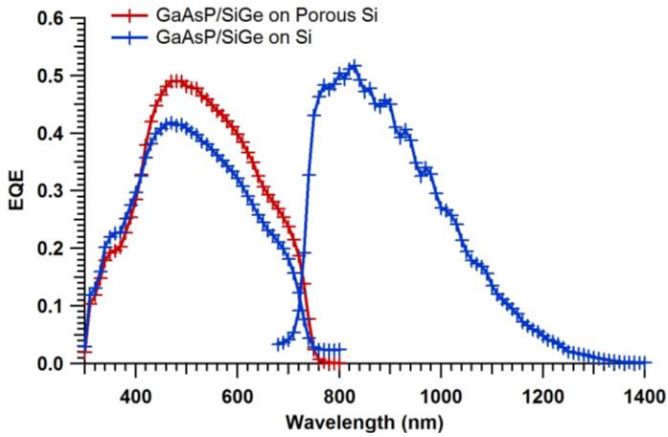

Fig. 5. EQE of the GaAsP/SiGe tandem cells without (in blue) and with porous silicon (in red).

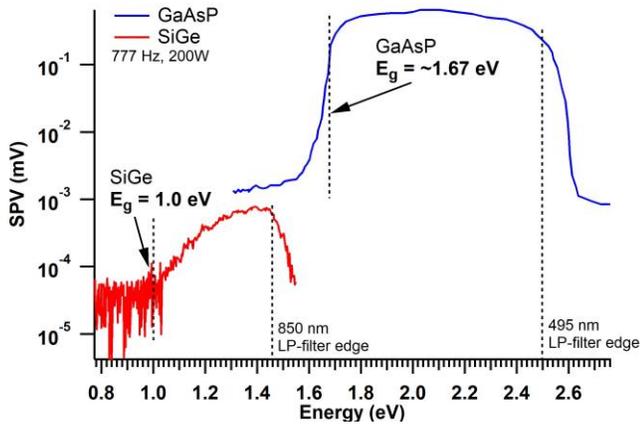

Fig. 6. SPV of the GaAsP top cell and the SiGe bottom cell where their approximate bandgaps are shown.

In order to cross-check if there is any photovoltaic activity in the SiGe bottom cell we carried out Spectral Photovoltage (SPV) measurements. SPV curves of the GaAsP top cell and SiGe bottom cell are shown in Fig. 6. This figure shows a strong response of the GaAsP cell (as expected) and a faint but measurable response from the SiGe bottom cell. In both cases, the upsurge of the voltage signal occurs at the energies that correspond to their bandgaps. As in first approximation the magnitude of SPV is proportional to the absorption coefficient, as far as charge separation does not depend on the incident wavelength, the standard analysis used for determining the values of direct and indirect bandgaps from absorption measurements can be also applied to SPV spectra in the energy range around the respective onsets. Results are indicated in Fig. 6 for both absorbers as reference, in excellent agreement with expected nominal values. These measurements confirm the development of a photovoltage in each subcell upon illumination and thus the existence of a tandem solar cell in this architecture, although with a severely shunted bottom cell.

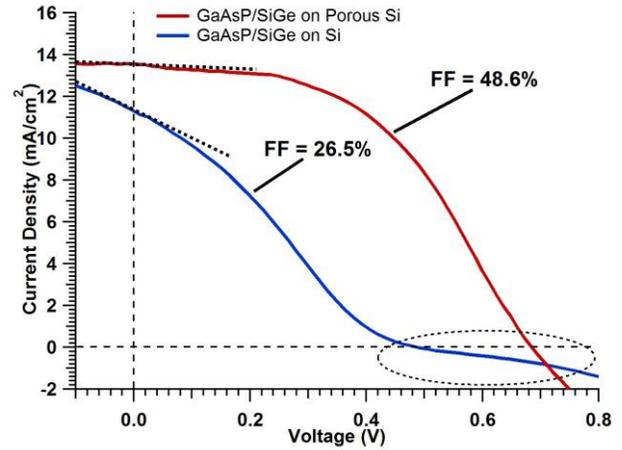

Fig. 7. I-V curves of the GaAsP/SiGe without (in blue) and with porous silicon (in red). It is worthy to note the gain in FF in the new structure.

One-sun I-V curves of two representative GaAsP/SiGe tandems are represented in Fig. 7, red for the design on porous silicon and blue for the design on conventional substrates. As can be observed, in both cases the short-circuit current densities measured are acceptable (the devices lack an ARC) and agree well with the differences observed in EQE. Looking at the shape of the curve beyond $V_{OC}$, the existence of a parasitic diode in reverse bias is also clearly noticeable, though less marked in the new structure. This diode is probably related to the tunnel junction or an adjacent layer. In conventional multijunction cells on Ge substrates, it has been reported that Ge can diffuse into the GaInP nucleation layer for over several hundred nm, causing a strong n-type background doping [28]. In our structure, we have an analogous situation in which a 100 nm n-GaInP nucleation layer is grown on a Ge cap layer (see Fig. 1). It is possible that Ge diffusion during the growth of the GaAsP top cell may reach the tunnel junction p-side –or its cladding layer–, reverting its polarity or, at least, greatly compensating its purportedly high doping. This situation would originate a diode in reverse polarity. The growth of a thicker GaInP nucleation layer could avoid this issue. For the same reason, the series resistance of the device cannot be assessed straightforwardly from the I-V curve due to the curvature caused by the parasitic diode. Porous silicon has been reported to cause high series resistance due to its lower conductivity [29]. However, the most remarkable feature is the limited FF and low $V_{OC}$ in both designs. In the case of tandems grown on conventional substrates, the low shunt resistance is very evident in the I-V causing deleterious effects in both FF and $V_{OC}$. The low $R_{shunt}$ is caused by a high crack density in the cell, which are both efficient recombination centers for minority carriers and act as electrical conduction paths producing a shunting behaviour. However, in samples grown on porous Si substrates, clear improvements in both *FF* (from 26.5% to 48.6%) and $V_{OC}$ are observable. Therefore, it can be concluded that the implementation of the porous silicon layer reduces the electrical shorting paths, thereby increasing the device performance, showing promise for further improvements.

## IV. Conclusions

Within the research line of the integration of III-V on silicon, we have developed a tandem GaAsP/SiGe solar cell grown on silicon through group-IV reverse graded buffer layers. In contrast to previous designs, the structure has been grown on substrates with a subsurface porous silicon layer. Although these wafers can reduce the threading dislocation density and allow the growth of thinner buffers, reverse graded buffer layers have shown to be prone to the appearance of cracks. The incorporation of the porous silicon has improved the GaAsP top cell and has repressed crack propagation, increasing the shunt resistance, which entails an improvement of the solar cell performance.

## Acknowledgements

This work was supported by the Spanish *Ministerio de Ciencia, Innovación y Universidades* through projects TEC2015-66722-R, RTI2018-094291-B-I00 and ENE2017-89561-C4-2-R; and by the *Comunidad de Madrid* through the project S2018/EMT-4308 (MADRID-PV2-CM). I. García acknowledges the financial support from his Ramón y Cajal grant (RYC-2014-15621).